\documentclass{ifacconf}

\usepackage{graphicx}      
\usepackage{natbib}        
\usepackage{color}
\usepackage{amsmath} 
\usepackage{amssymb}  
\usepackage{multicol}
\usepackage{enumitem}   
\usepackage{epstopdf}

\newcommand{\norm}[1]{\left\lVert#1\right\rVert}
\newcommand{\qedsymbol}{\rule{1.5ex}{1.5ex}}

\begin{document}
\begin{frontmatter}

\title{Robust integral action of port-Hamiltonian systems} 


\author[First]{Joel Ferguson} 
\author[Second]{Alejandro Donaire} 
\author[Third]{Romeo Ortega}
\author[First]{Richard H. Middleton}

\address[First]{School of Electrical Engineering and Computing and PRC CDSC, The University of Newcastle, Callaghan, NSW 2308, Australia (e-mail: Joel.Ferguson@uon.edu.au, Richard.Middleton@newcastle.edu.au).}
\address[Second]{Department of Electrical Engineering and Information Theory and PRISMA Lab, University of Naples Federico II, Napoli 80125, Italy, and with the School of Electrical Eng. and Comp. Sc. of the Queensland University of Technology, Brisbane, QLD, Australia (e-mail: Alejandro.Donaire@unina.it).}
\address[Third]{Laboratoire des Signaux et Syst\`emes, CNRS-SUPELEC, 91192, Gif-sur-Yvette, France (e-mail: romeo.ortega@lss.supelec.fr).}

\begin{abstract}                

Interconnection and damping assignment, passivity-based control (IDA-PBC) has proven to be a successful control technique for the stabilisation of many nonlinear systems. In this paper, we propose a method to robustify a system which has been stabilised using IDA-PBC with respect to constant, matched disturbances via the addition of integral action. The proposed controller extends previous work on the topic by being robust against the damping of the system, a quantity which may not be known in many applications.

\end{abstract}


\end{frontmatter}

\section{Introduction}
Interconnection and damping assignment, passivity-based control (IDA-PBC) is a design methodology for nonlinear systems whereby the closed-loop system is designed to be port-Hamiltonian (pH) \citep{Ortega2004}. This technique has been successfully applied to electrical systems \citep{Gonzalez2008}, mechanical systems \citep{Donaire2016a} and fluid systems \citep{Ortega2008}, amongst others. Integral action can be applied to systems stabilised using IDA-PBC to robustify the closed-loop with respect to constant disturbances. The process of adding integral action has been discussed by \cite{Donaire2009}; \cite{Ortega2012}; \cite{Romero2013a}; \cite{Ferguson2015}; \cite{Donaire2016}; \cite{Ferguson}; \cite{2017arXiv171006070F}.

As noted by \cite{Gomez-Estern2004}, application of IDA-PBC often neglects to account for the presence of physical damping within a system. In some cases, such as fully actuated mechanical systems, the presence of physical damping can be beneficial to the stability of the closed-loop systems as the damping preserves the passivity properties of the closed-loop. However, in other cases, physical damping has the ability to inject energy into the closed-loop, destroying the passivity properties of the system. This situation was explored by \cite{Gomez-Estern2004} and necessary and sufficient conditions for the existence of a stabilising control law in the presence of physical damping were provided. Similar problems have been investigated by \cite{Woolsey2004} using controlled Lagrangians and \cite{Romero2015} who developed a speed observer for pH systems perturbed by constant disturbances and with unknown damping is designed. Even in the case that the closed-loop system with physical damping is shown to be stable, only a conservative estimate of the damping may be known (See Section 3.2 of \citep{Gomez-Estern2004}).

Unfortunately, previous approaches to add integral action to pH systems have required the damping structure of the system to be known. As the damping may not be known in many applications, integral action is often unable to be applied. In this work, we propose an extension to the control law proposed in \cite{2017arXiv171006070F} that does not require knowledge of the damping matrix. Rather, the scheme presented here assumes that there exists a constant upper bound of the part of the damping matrix related to the actuated state. The scheme is shown to preserve a desired equilibrium and rejecting the effects of a constant matched disturbance. Furthermore, it is then shown that if the output of the undisturbed open-loop system is detectable, the closed-loop system is detectable also. Consequently, this integral action scheme can be applied to systems that have been asymptotically stabilised using IDA-PBC whilst preserving the asymptotic stability properties of the original control design.

\noindent {\bf Notation.} In this paper we use the following notation: Let $x \in\mathbb{R}^n$, $x_a\in\mathbb{R}^m$, $x_u\in\mathbb{R}^s$. For real valued functions $\mathcal{H}(x)$, $\nabla\mathcal{H}\triangleq \left(\frac{\partial \mathcal{H}}{\partial x}\right)^\top$. For functions $\mathcal{G}(x_a,x_u)\in\mathbb{R}$, $\nabla_{x_i}\mathcal{G}\triangleq \left(\frac{\partial \mathcal{G}}{\partial x_i}\right)^\top$ where $i \in\{a,u\}$. $\norm{x}_{K}^2 = x^\top Kx$ where $K\in\mathbb{R}^{n\times n}$. $0_{m\times s}$ denotes a $m\times s$ matrix where each entry is 0.

\section{Problem formulation}\label{probform}
In this paper, a class systems that have been stabilised using IDA-PBC are considered. The dynamics of such systems are of the form:
\begin{equation}\label{phdist}
	\begin{split}
		\begin{bmatrix}
		\dot x_a \\
		\dot x_u
		\end{bmatrix}
		&= 
		\begin{bmatrix}
		J_{aa}(x) - R_{aa}(x) & J_{au}(x) - R_{au}(x) \\
		- J_{au}^\top(x) - R_{au}^\top(x) & J_{uu}(x) - R_{uu}(x)
		\end{bmatrix}
		\begin{bmatrix}
		\nabla_{x_a}\mathcal{H} \\
		\nabla_{x_u} \mathcal{H}
		\end{bmatrix} \\
		&\phantom{---}+
		\begin{bmatrix}
		I_{m\times m} \\ 0_{s\times m}
		\end{bmatrix}
		(u-d) \\
		y
		&=
		\nabla_{x_a}\mathcal{H},
	\end{split}
\end{equation}
where $x = \operatorname{col}(x_a, x_u)\in\mathbb{R}^n$ is the state vector, $x_a \in \mathbb{R}^m$ and $x_u \in \mathbb{R}^s$ where $s \triangleq n-m$. The interconnection and damping matrices satisfy 
\begin{equation}
	\begin{split}
		J(x) = \begin{bmatrix} J_{aa}(x) & J_{au}(x) \\ -J_{au}^\top(x) & J_{uu}(x) \end{bmatrix} &= -J^\top(x) \\
		R(x) = \begin{bmatrix} R_{aa}(x) & R_{au}(x) \\ R_{au}^\top(x) & R_{uu}(x) \end{bmatrix} &= R^\top(x)  \geq 0
	\end{split}
\end{equation}
respectively,
the function $\mathcal{H}:\mathbb{R}^n \to \mathbb{R}$ is the Hamiltonian that represents the total energy of the system, $u\in\mathbb{R}^m$ is the input, $y \in \mathbb{R}^m$ is the output and $d \in \mathbb{R}^m$ is a constant disturbance.
We assume that the system \eqref{phdist} has an isolated equilibrium $x^\star = (x_a^\star,x_u^\star)$ that is a minimum of the Hamiltonian, which ensures that $x^\star$ is a stable equilibrium of \eqref{phdist} when $u = d = 0_{m\times 1}$.

\textbf{Problem statement:}
	Given the pH system \eqref{phdist}, design a dynamic control law $u=u(x_a,x_u,x_c)$, where $x_c \in \mathbb{R}^{m}$ is the state of the controller, such that:
	\begin{itemize}
	\item the closed-loop has an (asymptotically) stable equilibrium at $(x,x_c) = (x^\star,x_c^\star)$ for some $x_c^\star\in\mathbb{R}^m$.
	\item the closed-loop dynamics can be written in the pH form.
	\item the control law is independent of $R(x)$.
	\end{itemize}

The main contribution of this work is the consideration that $R(x)$ is an unknown, possibly state dependant, quantity. This may be the case for many practical control systems such as fully actuated mechanical systems stabilised using potential energy shaping or pH systems that have been made passive using the techniques outlined in \citep{Gomez-Estern2004}.

\section{Previous work}
In this section, the integral action scheme of \cite{2017arXiv171006070F} is reviewed. In the case that $R(x)$ is known, the following control law was proposed as a solution to the integral action problem:
\begin{eqnarray}\label{OldcontrolLaw}
		u &=& (-J_{aa}+R_{aa}+J_{c_1}-R_{c_1}-R_{c_2})\nabla_{x_a} \mathcal{H} \nonumber \\ 
		&&\phantom{--} + (J_{c_1}-R_{c_1})K_i(x_a-x_c) + 2 R_{au} \nabla_{x_u} \mathcal{H} \nonumber \\
		\dot{x}_c&=&-R_{c_2}\nabla_{x_a}\mathcal{H} + (J_{au} + R_{au}) \nabla_{x_u}\mathcal{H},
\end{eqnarray}
where $x_c \in \mathbb{R}^m$ is the state of the controller and the matrices $K_i, J_{c_1}, R_{c_1}, R_{c_2} \in \mathbb{R}^{m\times m}$ are free to be chosen such that $K_i > 0$, $J_{c_1}=-J_{c_1}^\top$, $R_{c_1} > 0$, $R_{c_2} \geq 0$. 

By defining the coordinates
\begin{equation}\label{wCoords}
	w
	=
	\begin{bmatrix}
		w_a \\ w_u \\ w_c
	\end{bmatrix}
	=
	\begin{bmatrix}
		x_a \\ x_u \\ x_a-x_c
	\end{bmatrix},
\end{equation}
the closed-loop dynamics can then be written as the disturbed pH system
\begin{equation}\label{iacl}
		\begin{bmatrix}
			\dot{w}_a \\
			\dot{w}_u \\
			\dot{w}_c
		\end{bmatrix}
		=
		\left[ J_{cl}(w) - R_{cl}(w) \right]
		\begin{bmatrix}
			\nabla_{w_a}\mathcal{H}_{cl} \\
			\nabla_{w_u}\mathcal{H}_{cl} \\
			\nabla_{w_c}\mathcal{H}_{cl}
		\end{bmatrix}
		-
		\begin{bmatrix}
			d \\ 0_{s\times 1} \\ d
		\end{bmatrix},
\end{equation}
where
\begin{equation}\label{Jcl}
J_{cl}(x)=  
\begin{bmatrix}
J_{c_1} & J_{au} + R_{au} & J_{c_1} \\
- J_{au}^\top - R_{au}^\top & J_{uu} & 0_{s\times m} \\
J_{c_1} & 0_{m\times s} & J_{c_1}
\end{bmatrix},
\end{equation}
and
\begin{equation}\label{Rcl}
R_{cl}(x)=  
\begin{bmatrix}
R_{c_1} + R_{c_2} & 0_{m\times s} & R_{c_1} \\
0_{s\times m} & R_{uu} & 0_{s\times m} \\
R_{c_1} & 0_{m\times s} & R_{c_1} 
\end{bmatrix}
\end{equation}
and $\mathcal{H}_{cl}:\mathbb{R}^{2m+n}\to\mathbb{R}$ is the closed-loop Hamiltonian defined as 
\begin{equation}\label{CLHamiltonian}
\mathcal{H}_{cl}(w_a,w_u,w_c) = \mathcal{H}(w_a,w_u) + \frac12\norm{w_c}_{K_i}^2.
\end{equation}
The function
\begin{equation}
	\mathcal{W}_l(w_a,w_u,w_c) = \mathcal{H}(w_a,w_u) + \frac12 \norm{w_c - K_i^{-1}R_{c_1}^{-1}d}_{K_i}^2
\end{equation}
is a Lyapunov function of the closed loop, which ensures stability of \eqref{iacl} (see \cite{2017arXiv171006070F} for a detailed discussion of this design).

\section{Main result}
The previously proposed control law \eqref{OldcontrolLaw} uses the terms $R_{aa}$ and $R_{au}$. Thus, it requires knowledge of the open-loop system damping in \eqref{phdist}. In this section, the control scheme \eqref{OldcontrolLaw} is modified such that is does not rely of the damping of the system, $R(x)$.

\begin{prop} 
Consider the system \eqref{phdist} in closed-loop with the controller 
\begin{equation}\label{controlLaw}
	\begin{split}
		u &= -(J_{aa}+D_{c_1}+D_{c_2}+D_{c_3})\nabla_{x_a}\mathcal{H} \\
		&\phantom{---} -(D_{c_1}+D_{c_3})K_i(x_a-x_c) \\
		\dot{x}_c&=-(D_{c_2}+D_{c_3})\nabla_{x_a}\mathcal{H} + J_{au} \nabla_{x_u}\mathcal{H},
	\end{split}
\end{equation}
where $x_c \in \mathbb{R}^m$ is the state of the controller and the matrices $D_{c_1}, D_{c_2}, D_{c_3} \in \mathbb{R}^{m\times m}$ are tuning parameters satisfying $D_{c_1} > 0$, $D_{c_2} > 0$, $D_{c_3} > 0$. The closed-loop dynamics are given by
\begin{equation}\label{IA:ClsdLoop}
	\begin{split}
		\begin{bmatrix}
			\dot{w}_a \\
			\dot{w}_u \\
			\dot{w}_c
		\end{bmatrix}
		&=  
		\left\lbrace
			\mathcal{J}(x)-\mathcal{R}(x)
		\right\rbrace
		\begin{bmatrix}
			\nabla_{w_a}\mathcal{W} \\
			\nabla_{w_u}\mathcal{W} \\
			\nabla_{w_c}\mathcal{W}
		\end{bmatrix},
	\end{split}
\end{equation}
with
\begin{equation}
	\begin{split}
		\mathcal{J}
		&=
		\begin{bmatrix}
			0_{m\times m} & J_{au} & 0_{m\times m} \\
			- J_{au}^\top & J_{uu} & 0_{s\times m} \\
			0_{m\times m} & 0_{m\times s} & 0_{m\times m}
		\end{bmatrix} \\
		\mathcal{R}
		&=
		\begin{bmatrix}
			R_{aa} + D_{c_1} + D_{c_2} + D_{c_3} & R_{au} & D_{c_1}+D_{c_3} \\
			R_{au}^\top & R_{uu} & 0_{s\times m} \\
			R_{aa} + D_{c_1} & R_{au} & D_{c_1} + D_{c_3}
		\end{bmatrix} \\
	\end{split}
\end{equation}
where $w$ is defined by \eqref{wCoords} and 
\begin{equation}
	\begin{split}
		\mathcal{W}(w_a,w_u,w_c)
		&=
		\mathcal{H}(w_a,w_u) \\
		&\phantom{-}+ \frac12\norm{w_c+K_i^{-1}(D_{c_1}+D_{c_3})^{-1}d}_{K_i}^2.
	\end{split}
\end{equation}
Furthermore, the closed-loop system \eqref{IA:ClsdLoop} has an equilibrium point
	\begin{equation}\label{EqPoint}
		w^\star = (x_a^\star,x_u^\star,-K_i^{-1}(D_{c_1}+D_{c_3})^{-1}d).
	\end{equation}
\end{prop}

\begin{pf} 
Substituting the control law \eqref{controlLaw} into the dynamics of $x_a$ in \eqref{phdist} results in
\begin{equation}\label{IA:robust:xa}
	\begin{split}
		\dot x_a &= -(R_{aa}+D_{c_1}+D_{c_2}+D_{c_3})\nabla_{x_a}\mathcal{H} + (J_{au} - R_{au}) \\
		&\phantom{---} \times \nabla_{x_u}\mathcal{H} - (D_{c_1}+D_{c_3})K_i(x_a - x_c) - d
	\end{split}
\end{equation}
Noticing that
\begin{equation}
	\nabla \mathcal{W}
	=
	\begin{bmatrix}
		\nabla_{x_a}\mathcal{H} \\\nabla_{x_u}\mathcal{H} \\ K_i (x_a - {x}_c) + (D_{c_1}+D_{c_3})^{-1}d
	\end{bmatrix},
\end{equation}
and substituting into \eqref{IA:robust:xa} yields
\begin{equation}
	\begin{split}
		\dot x_a &= -(R_{aa}+D_{c_1}+D_{c_2}+D_{c_3})\nabla_{w_a}\mathcal{W} + (J_{au} - R_{au})  \\
		&\phantom{---} \times \nabla_{w_u}\mathcal{W} - (D_{c_1}+D_{c_3})\nabla_{w_c}\mathcal{W},
	\end{split}
\end{equation}
which is the dynamics of $w_a$ in  \eqref{IA:ClsdLoop}. 

The proof follows noticing that the second rows of \eqref{phdist} and \eqref{IA:ClsdLoop} match. Finally, writing the controller dynamics \eqref{controlLaw} as
\begin{equation}
	\dot{x}_c=-(D_{c_2}+D_{c_3})\nabla_{w_a}\mathcal{W} + J_{au} \nabla_{w_u}\mathcal{W}
\end{equation}
and subtracting it from the expression of $\dot x_a$ above yields the last row of \eqref{IA:ClsdLoop}. Thus, the closed-loop dynamics can be written in the pH form \eqref{IA:ClsdLoop} as claimed.

To verify that \eqref{EqPoint} is an equilibrium of the closed-loop system, note that $\mathcal{W}$ is minimised by \eqref{EqPoint} and satisfies $\nabla \mathcal{W}|_{w=w^\star} = 0_{(2m+s)\times 1}$. Substituting this equilibrium gradient into \eqref{IA:ClsdLoop} verifies that \eqref{EqPoint} is indeed an equilibrium point.
\hfill\qedsymbol
\end{pf}

When designing control laws with IDA-PBC, it is typically insisted that $\mathcal{R}$ be symmetric matrix which is not the case in this work. Despite this, it will be shown in subsequent sections that $\mathcal{R}$ satisfies $\nabla^\top \mathcal{W} \; \mathcal{R} \;  \nabla \mathcal{W} \geq 0$, and therefore it is a dissipative term that help to stabilise the closed loop. Similar class of terms have been used in \citep{Batlle2009} and \citep{Donaire2016b} in the context of simultaneous IDA-PBC for induction motor and mechanical systems, respectively.



In order to modify the control scheme to be robust against the damping of the system, we require a constant upper bound for $R_{aa}(x)$.

\begin{assum}\label{AssDampingEst}
	There exists a constant matrix $D_{c_3}\in\mathbb{R}^{m\times m}$ with $D_{c_3} > 0$ such that 
	\begin{equation}
		3D_{c_3} - R_{aa}(x) > 0.
	\end{equation}
\end{assum}
Provided that the open-loop system satisfies Assumption \ref{AssDampingEst}, the control law \eqref{controlLaw} solves the integral action problem.

\begin{prop}\label{PropStability}
Under Assumption~\ref{AssDampingEst},	the closed-loop system \eqref{IA:ClsdLoop} satisfies:
	\begin{enumerate}[label=(\roman*)]
	\item The equilibrium \eqref{EqPoint} is a stable. \label{Stability}
	\item If the output
		\begin{equation}\label{DetectOut}
			y_d
			=
			\begin{bmatrix}
				\nabla_{x_a}\mathcal{H} \\
				K_i(x_a-x_c) + (D_{c_1}+D_{c_3})^{-1}d
			\end{bmatrix}
		\end{equation}
		 is detectable, the equilibrium \eqref{EqPoint} is asymptotically stable. \label{AsymptStability}
	\item In addition, if $\mathcal{H}$ is radially unbounded, the stability properties are global. \label{GlobStability}
	\end{enumerate}
\end{prop}

\begin{pf}
	To verify \ref{Stability}, notice that $\mathcal{W}$ is positive definite in some neighbourhood of \eqref{EqPoint}.
	Taking $\mathcal{W}$ as a Lyapunov candidate for the closed-loop system, the time derivative of $\mathcal{W}$ satisfies
	\begin{equation}\label{WTimeDer}
		\begin{split}
			\dot{\mathcal{W}}
			&=
			\begin{bmatrix}
				\nabla_{w_a}\mathcal{W} \\
				\nabla_{w_u}\mathcal{W} \\
				\nabla_{w_c}\mathcal{W}
			\end{bmatrix}^\top
			\left\lbrace
				\mathcal{J}-\mathcal{R}
			\right\rbrace
			\begin{bmatrix}
				\nabla_{w_a}\mathcal{W} \\
				\nabla_{w_u}\mathcal{W} \\
				\nabla_{w_c}\mathcal{W}
			\end{bmatrix} \\
			&=
			-
			\begin{bmatrix}
				\nabla_{w_a}\mathcal{W} \\
				\nabla_{w_u}\mathcal{W} \\
				\nabla_{w_c}\mathcal{W}
			\end{bmatrix}^\top
			\mathcal{R}
			\begin{bmatrix}
				\nabla_{w_a}\mathcal{W} \\
				\nabla_{w_u}\mathcal{W} \\
				\nabla_{w_c}\mathcal{W}
			\end{bmatrix}.
		\end{split}
	\end{equation}
	Clearly, stability of the closed-loop system is established if $\mathcal{R}$ is positive semi-definite. To show this property to be true, $\mathcal{R}$ is broken into two parts:
	\begin{equation}
		\begin{split}
			\mathcal{R}_1
			&=
			\begin{bmatrix}
				R_{aa} + D_{c_3} & R_{au} & D_{c_3} \\
				R_{au}^\top & R_{uu} & 0_{s\times m} \\
				R_{aa} & R_{au} & D_{c_3}
			\end{bmatrix} \\
			\mathcal{R}_2
			&=
			\begin{bmatrix}
				D_{c_1} + D_{c_2} & 0_{m\times s} & D_{c_1} \\
				0_{s\times m} & 0_{s\times s} & 0_{s\times m} \\
				D_{c_1} & 0_{m\times s} & D_{c_1}
			\end{bmatrix}.
		\end{split}
	\end{equation}
	Clearly, $\mathcal{R} = \mathcal{R}_1+\mathcal{R}_2$.
	
	Now $\mathcal{R}_1$ is shown to be positive semi-definite. The symmetric part of $\mathcal{R}_1$ is given by
	\begin{equation}
		\frac12(\mathcal{R}_1+\mathcal{R}_1^\top)
		=
		\begin{bmatrix}
			R_{aa} + D_{c_3} & R_{au} & \frac12(R_{aa}+D_{c_3}) \\
			R_{au}^\top & R_{uu} & \frac12 R_{au}^\top \\
			\frac12(R_{aa}+D_{c_3}) & \frac12 R_{au} & D_{c_3}
		\end{bmatrix}.
	\end{equation}
	The term $R_{aa} + D_{c_3}$ is positive definite as $R\geq 0$ and $D_{c_3} > 0$. Thus, the Schur complement condition for positive semi-definiteness states that $\frac12(\mathcal{R}_1+\mathcal{R}_1^\top)$ is positive semi-definite if and only if
	\begin{equation}\label{D1:1}
		\begin{split}
			&-
			\begin{bmatrix}
				R_{au}^\top \\
				\frac12(R_{aa}+D_{c_3})
			\end{bmatrix}
			(R_{aa} + D_{c_3})^{-1}
			\begin{bmatrix}
				R_{au} & \frac12(R_{aa}+D_{c_3}) \\
			\end{bmatrix} \\
			&\phantom{---------------}+
			\begin{bmatrix}
				R_{uu} & \frac12 R_{au}^\top \\
				\frac12 R_{au} & D_{c_3}
			\end{bmatrix}
			\geq 0.
		\end{split}
	\end{equation}
	Expanding \eqref{D1:1} results in
	\begin{equation}\label{D1:2}
		\begin{split}
			&
			\begin{bmatrix}
				R_{uu} & \frac12 R_{au}^\top \\
				\frac12 R_{au} & D_{c_3}
			\end{bmatrix}
			-
			\begin{bmatrix}
				R_{au}^\top(R_{aa} + D_{c_3})^{-1}R_{au} & \frac12 R_{au}^\top \\
				\frac12 R_{au} & \frac14(R_{aa}+D_{c_3})
			\end{bmatrix} \\
			&\phantom{---}=
			\begin{bmatrix}
				\mathbf{D}_1 & 0_{s\times m} \\
				0_{m\times s} & \mathbf{D}_2
			\end{bmatrix}
			\geq 0,
		\end{split}
	\end{equation}
	where $\mathbf{D}_1 = R_{uu} - R_{au}^\top(R_{aa} + D_{c_3})^{-1}R_{au}$ and $\mathbf{D}_2 = D_{c_3} - \frac14(R_{aa}+D_{c_3})$. Condition \eqref{D1:2} is true if both $\mathbf{D}_1 \geq 0$ and $\mathbf{D}_2 \geq 0$.
	
	To see that $\mathbf{D}_1 \geq 0$, notice that the matrix
	\begin{equation}
		\begin{split}
			\begin{bmatrix}
			R_{aa} + D_{c_3} & R_{au} \\
			R_{au}^\top & R_{uu}
			\end{bmatrix}
		\end{split}
	\end{equation}
	is positive semi-definite and $R_{aa} + D_{c_3} > 0$. Thus, by application of the Schur complement condition for positive semi-definiteness, $\mathbf{D}_1 \geq 0$. $\mathbf{D}_2$ is positive semi-definite as $D_{c_3}$ satisfies Assumption \ref{AssDampingEst}. Thus, $\mathcal{R}_1$ is positive semi-definite.
	
	Using the fact that $\frac12(\mathcal{R}_1+\mathcal{R}_1^\top) \geq 0$, \eqref{WTimeDer} can be simplified to
	\begin{equation}
		\begin{split}
			\dot{\mathcal{W}}
			&=
			-
			\begin{bmatrix}
				\nabla_{w_a}\mathcal{W} \\
				\nabla_{w_u}\mathcal{W} \\
				\nabla_{w_c}\mathcal{W}
			\end{bmatrix}^\top
			(\mathcal{R}_1+\mathcal{R}_2)
			\begin{bmatrix}
				\nabla_{w_a}\mathcal{W} \\
				\nabla_{w_u}\mathcal{W} \\
				\nabla_{w_c}\mathcal{W}
			\end{bmatrix} \\
			&\leq
			-
			\begin{bmatrix}
				\nabla_{w_a}\mathcal{W} \\
				\nabla_{w_u}\mathcal{W} \\
				\nabla_{w_c}\mathcal{W}
			\end{bmatrix}^\top
			\underbrace{
			\begin{bmatrix}
				D_{c_1} + D_{c_2} & 0_{m\times s} & D_{c_1} \\
				0_{s\times m} & 0_{s\times s} & 0_{s\times m} \\
				D_{c_1} & 0_{m\times s} & D_{c_1}
			\end{bmatrix}}_{\mathcal{R}_2}
			\begin{bmatrix}
				\nabla_{w_a}\mathcal{W} \\
				\nabla_{w_u}\mathcal{W} \\
				\nabla_{w_c}\mathcal{W}
			\end{bmatrix} \\
			&\leq
			-\norm{\nabla_{w_a}\mathcal{W}}_{D_{c_2}}^2
			-\norm{\nabla_{w_a}\mathcal{W}+\nabla_{w_c}\mathcal{W}}_{D_{c_1}}^2
		\end{split}
	\end{equation}
	proving stability. Claim \ref{AsymptStability} follows by invoking LaSalle's invariance principle. Finally, to verify \ref{GlobStability}, notice that if $\mathcal{H}$ is radially unbounded in $x_a, x_u$, $\mathcal{W}$ is radially unbounded in $w_a, w_u, w_c$, implying that the stability results are global.
	\hfill\qedsymbol
	\end{pf}
	The closed-loop stable by Proposition \ref{PropStability}. In the following corollary it is shown that if the output of the undisturbed open-loop system is detectable, the closed-loop system is asymptotically stable.
	\begin{cor}
		If the output $y$ of the undisturbed open-loop system \eqref{phdist} is detectable when $u = d = 0_{m\times 1}$, then the equilibrium point \eqref{EqPoint} of the disturbed closed-loop system \eqref{IA:ClsdLoop} is asymptotically stable.
	\end{cor}
	\begin{pf}
		By Proposition \ref{PropStability}, the closed-loop system is asymptotically stable if the output $y_d$ is detectable. To verify this, first restrict the control law to the set defined by $y_d = 0_{m\times 1}$:
		\begin{equation}
			\begin{split}
				u|_{y_d = 0_{m\times 1}} &= d \\
				\dot{x}_c|_{y_d = 0_{m\times 1}} &= J_{au} \nabla_{x_u}\mathcal{H},
			\end{split}
		\end{equation}
		Substituting into \eqref{phdist} and noting the $y|_{y_d = 0_{m\times 1}} = 0_{m\times 1}$ recovers the zero dynamics of \eqref{phdist} when $u = d = 0_{m\times 1}$.
		\hfill\qedsymbol
	\end{pf}

\section{2-degree of freedom manipulator}
The 2-degree of freedom planar manipulator in closed-loop with energy shaping control has the dynamic equations
\begin{equation}\label{ex:2dofManip}
	\begin{split}
		\begin{bmatrix}
			\dot p \\ \dot q
		\end{bmatrix}
		&=
		\begin{bmatrix}
			-K_d-D_p(q,\dot q) & -I_{2\times 2} \\
			I_{2\times 2} & 0_{2\times 2}
		\end{bmatrix}
		\begin{bmatrix}
			\nabla_p \mathcal{H}_d \\
			\nabla_q \mathcal{H}_d
		\end{bmatrix} \\
		&\phantom{---}+
		\begin{bmatrix}
			I_{2\times 2} \\
			0_{2\times 2}
		\end{bmatrix}
		(u-d) \\
		y
		&=
		\nabla_p \mathcal{H}_d \\
		\mathcal{H}_d(q,p)
		&=
		\frac12 p^\top M^{-1}(q)p + \frac12(q-q_d)^\top K_p(q-q_d)
	\end{split}
\end{equation}
where $q = (\theta_1,\theta_2)$ are the angular rotations of the first and second manipulator links respectively, $q_d\in\mathbb{R}^2$ is the target configuration, the momentum is defined by $p = M(q)\dot q$,
\begin{equation}
	M(q)
	=
	\begin{bmatrix}
		a_1+a_2+2b\cos\theta_2 & a_2+b\cos\theta_2 \\
		a_2+b\cos\theta_2 & a_2
	\end{bmatrix},
\end{equation}
$a_1, a_2, b$ are constant system parameters, $K_d,K_p\in \mathbb{R}^{2\times 2}$ are positive definite terms associated with damping injection and energy shaping, respectively, and $d\in\mathbb{R}^2$ is a constant disturbance \citep{Dirksz2012a}. Here, we consider an additional term $D_p\in \mathbb{R}^{2\times 2}$ which contain the physical damping terms of the open-loop system. Assuming that the only physical damping within the system occurs at the joints, $D_p$ has the form
\begin{equation}
	D_p(q,\dot q)
	=
	\begin{bmatrix}
		-d_1(q,\dot q)-d_2(q,\dot q) & d_2(q,\dot q) \\
		d_2(q,\dot q) & -d_2(q,\dot q)
	\end{bmatrix}
\end{equation}
where $d_1(q,\dot q)$ and $d_2(q,\dot q)$ are the friction characteristics of the first and second joins respectively.

The system \eqref{ex:2dofManip} is of the form \eqref{phdist} with $x_a = p, x_u = q$ and $R_{aa} = K_d+D_p$. Provided that the terms $d_1(q,\dot q), d_2(q,\dot q)$ can be upper bounded, Assumption \ref{AssDampingEst} is satisfied and the integral action controller \eqref{controlLaw} can be used for disturbance rejection.

For the purpose of simulation, the physical damping terms were taken to be $d_1(\dot q) = \frac{\beta_1}{\sqrt{\alpha_1+\dot q_1^2}}$, $d_2(\dot q) = \frac{\beta_2}{\sqrt{\alpha_2+(\dot q_1-\dot q_2)^2}}$ which are approximations of Coulomb friction at the pivot points \citep{Gomez-Estern2004}. The system was simulated using the following parameters: $a_1 = 0.1476$, $a_2 = 0.0725$, $b = 0.0858$, $\alpha_1 = \alpha_2 = 0.1$, $\beta_1 = \beta_2 = 2$. The controller parameters were chosen to be $K_d = \operatorname{diag}(5,5)$, $K_p = \operatorname{diag}(30,20)$, $K_i = D_{c_1} = D_{c_2} = I_{2\times 2}$, $D_{c_3} = \frac12(K_d + \bar D_p)$,
\begin{equation}
	\bar D_p
	=
	\begin{bmatrix}
		-\bar d_1 - \bar d_2 & \bar d_2 \\
		\bar d_2 & -\bar d_2
	\end{bmatrix}
\end{equation}
and $\bar d_1 = \frac{\alpha_1}{\beta_1}$, $\bar d_2 = \frac{\alpha_2}{\beta_2}$.

\begin{figure}
\begin{center}
\includegraphics[width=9cm]{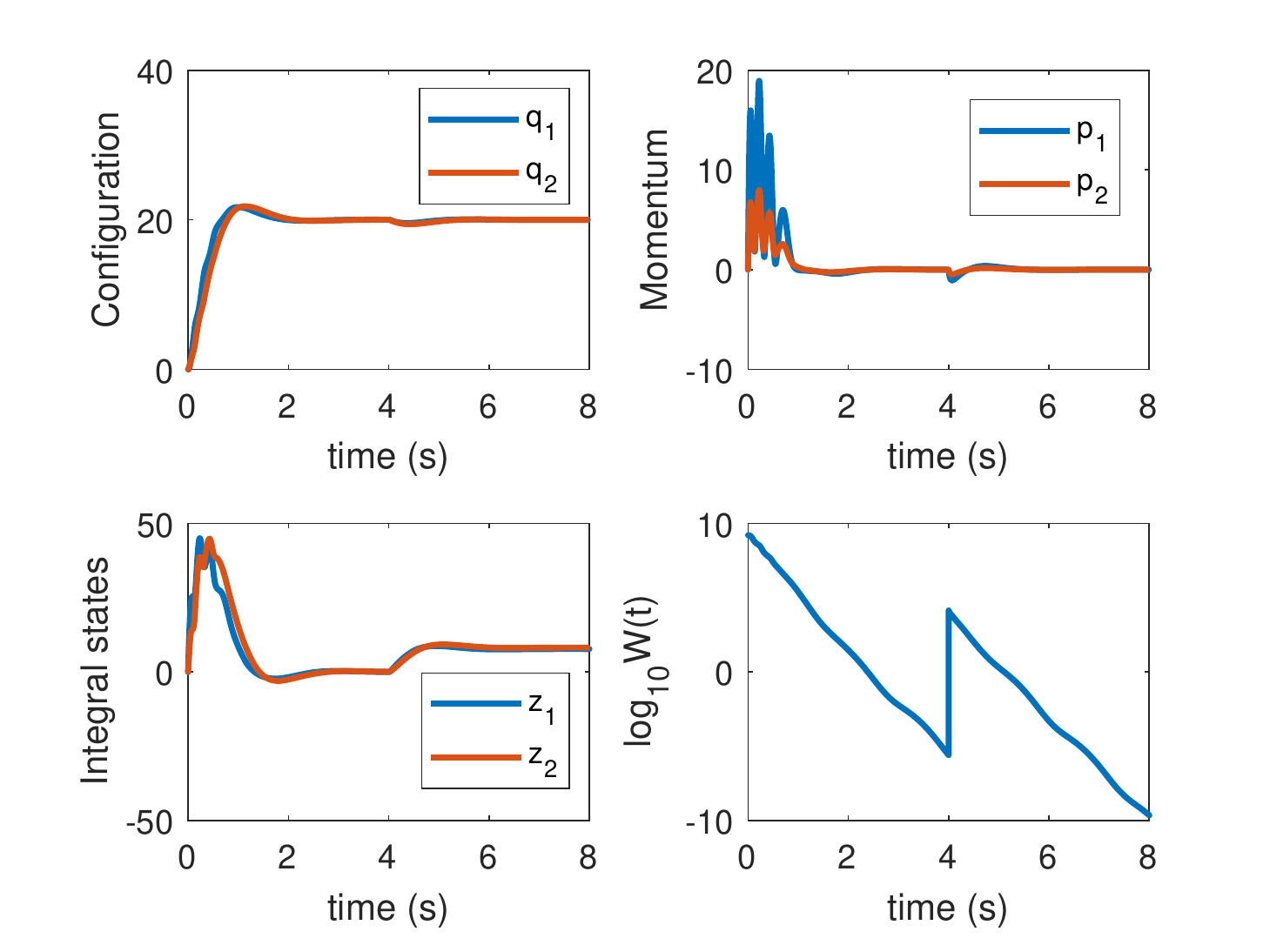}    
\caption{The 2 degree of freedom manipulator in closed-loop with an energy shaping control law and integral action. The system is initially specified to drive to the configuration $(20,20)$. At time $t=4$, a matched disturbance of $d = (50,30)$ is applied to the system. The integral action controller rejects to effect of the disturbance from the closed-loop system.} 
\label{2DOFfig1}
\end{center}
\end{figure}

The system was initialised at $q(0) = (0,0)$, $p(0) = (0,0)$, $x_c(0) = (0,0)$ and the target configuration set to $q_d = (20,20)$. At time $t=4$, the system was subjected to a matched disturbance of $d = (50,30)$.
Figure \ref{2DOFfig1} shows that the closed-loop system asymptotically converges to the desired final position, rejecting the effects of the unknown disturbance. 

\section{Conclusion}

In this paper, a method of applying integral action to systems that have been stabilised using IDA-PBC was presented. The method extends previous work on the topic by not requiring exact knowledge of the damping structure of the open-loop system. Rather, the control scheme utilises an upper bound on the damping structure. The integral action scheme was applied to the 2-degree of freedom planar manipulator and demonstrated in simulation to reject the effects of a constant matched disturbance.



\end{document}